\documentclass[preprint, prb,showkeys,amsmath,amssymb,longbibliography,citeautoscript]{revtex4-2}

\usepackage{amsmath}
\usepackage{amssymb}
\usepackage[hidelinks]{hyperref}
\usepackage{chemformula}
\usepackage{siunitx}    
\usepackage[capitalise]{cleveref}

\usepackage{graphicx}
\usepackage{dcolumn}
\usepackage{bm}
\usetikzlibrary{positioning}
\usepackage{siunitx}
\usepackage[version=3]{mhchem}
\DeclareSIUnit\angstrom{\text {Å}}
\usepackage{mathtools}

\DeclarePairedDelimiter{\abs}{\lvert}{\rvert} 
\setcitestyle{super}

\begin{document}

\title{Accurate Description of Ion Migration in Solid-State Ion Conductors from Machine-Learning Molecular Dynamics}

\author{Takeru Miyagawa}
\author{Namita Krishnan}
\author{Manuel Grumet}
\author{Christian~Rever\'{o}n~Baecker}
\author{Waldemar Kaiser}
\email{waldemar.kaiser@tum.de}
\author{David A. Egger}
\email{david.egger@tum.de}
\affiliation{%
  Department of Physics, TUM School of Natural Sciences, Technical University of Munich, 
  85748 Garching, Germany
}

\begin{abstract}
Solid-state ion conductors (SSICs) have emerged as a promising material class for electrochemical storage devices and novel compounds of this kind are continuously being discovered. High-throughout approaches that enable a rapid screening among the plethora of candidate SSIC compounds have been essential in this quest. 
While first-principles methods are routinely exploited in this context to provide atomic-level details on ion migration mechanisms, dynamic calculations of this type are computationally expensive and limit us in the time- and length-scales accessible during the simulations.
Here, we explore the potential of recently developed machine-learning force fields for predicting different ion migration mechanisms in SSICs. Specifically, we systematically investigate three classes of SSICs that all exhibit complex ion dynamics including vibrational anharmonicities: AgI, a strongly disordered Ag$^+$-conductor; Na$_3$SbS$_4$, a Na$^+$ vacancy conductor; and Li$_{10}$GeP$_2$S$_{12}$, which features concerted Li$^+$ migration. Through systematic comparison with \textit{ab initio} molecular dynamics data, we demonstrate that machine-learning molecular dynamics provides very accurate predictions of the structural and vibrational properties including the complex anharmonic dynamics in these SSICs. The \textit{ab initio} accuracy of machine-learning molecular dynamics simulations at relatively low computational cost open a promising path toward the rapid design of novel SSICs.

\end{abstract}

\maketitle
\pagebreak

\section{Introduction}
\label{sec:intro}

Solid-state ion conductors (SSICs) are essential components of electrochemical energy storage devices such as all-solid-state batteries \cite{PARK2010, Quartarone_2011, Bruce2012, zhou2017rechargeable, Ye_2020_intro, balaish2021processing, kim2021}. Compared to their liquid counterparts, SSICs offer improved safety and high energy density \cite{Lotsch2017, Placke_2017,guo2022solid}. Recent advances have increased the ion conductivities to values comparable to the ones of liquid electrolytes, making them a feasible alternative for future battery technologies \cite{zhang2018new, usiskin2021fundamentals}. The synergy among theoretical and experimental materials studies has led to a continuous improvement of our understanding of ion-conduction phenomena, thereby generating a plethora of novel material candidates for usage of SSICs in battery devices.

Deriving structure--property relations between SSIC compositions and their ionic conductivities is key when establishing design rules for screening potential SSIC candidates in  experimental and theoretical high-throughput approaches \cite{Rickert_1978, wang2015design, Kyuki2020, Kim_2021_design, Yang_2022}. Probing the ion dynamics within the solid host lattice experimentally can be achieved via impedance spectroscopy \cite{macdonald1987impedance,vadhva2021electrochemical} and scattering techniques, \textit{e.g.} Raman \cite{geisel1977liquid, langer2019present} or inelastic neutron scattering \cite{shull1995early, hudson2006vibrational}. The former technique provides important insights on the frequency range at which different ionic species migrate as well as on their densities. The latter ones complement this in important ways by providing the overall vibrational spectrum of the material, including the features associated with mobile cations and the host lattice. The interpretation of such spectra, however, is not always straightforward and often relies on suitable reference materials. Consequently, experimental efforts alone can fall short in providing a complete microscopic picture of the ion conduction mechanism.

Therefore, atomistic simulations are essential in complementing experimental investigations as they offer valuable insights into microscopic details of the conduction mechanism of mobile cations through the complex host lattice. Furthermore, they allow the virtual investigation of novel material candidates and structural phases to guide the experimental design of SSICs in a synergistic approach. \textit{Ab initio} molecular dynamics (AIMD) simulations have been widely used to study the ion dynamics and particularly the coupling between mobile ions and the host lattice in SSICs \cite{famprikis2019new, Sagotra_2019, Brenner_2020, Ding_2020,gupta2021fast,krenzer2022anharmonic}, unraveling a variety of interesting phenomena that drive the conduction of mobile cations. 
Such computational studies have suggested various interesting phenomena, including the 'paddle-wheel effect' that is coupling cation translational and polyanion rotational motions to impact cation conduction \cite{Jansen_1991, Stephen;Hull_2004,  zhang2019coupled, Xu_2022, zhang2022exploiting}. Furthermore, various SSICs were found to exhibit anharmonic host lattice dynamics, which involve the mobile ions being driven into strongly anharmonic regions of the potential energy surface \cite{Brenner_2020, zhang2020targeting, brenner2022anharmonic,ren2023extreme}. Moreover, the collective motion of cations through the soft lattice, in the form of concerted migration, was found to substantially alter the local potential barriers experienced by cations for a variety of lithium and sodium ion conductors \cite{jalem2013concerted,He2017,peng2020fast, zou2020relationships}. 

Despite being a powerful and useful tool to elucidate the dynamical properties of SSICs, it is well-known that AIMD simulations suffer from a tremendous computational cost associated with the need to perform density functional theory (DFT) calculations at each time step. This limits accessible simulation times to the picosecond range and further restricts us to a small number of material systems and properties at small length scales.
The prospect of machine-learning molecular dynamics (MLMD) is to overcome these shortcomings via much faster but equally accurate calculations of dynamical properties of materials \cite{unke2021machine,fedik2022extending}. More rapid computational schemes will not only accelerate the \textit{in silico} design of novel SSICs but in fact enable us to conduct more realistic calculations, e.g., regarding doped supercells. We anticipate that this will bring computational studies closer to laboratory realities and enhance their synergy with experimental synthesis and characterization \cite{butler2018machine,himanen2019data}.

Throughout an AIMD simulation, an enormous set of \textit{ab initio} forces and energies is collected for atoms that typically move within a relatively narrow phase space around their equilibrium positions. This scenario appears to be ideal for machine-learning force fields (MLFFs), i.e., predicting the potential energy surface around individual atoms via supervised machine-learning (ML) techniques. Various MLMD methods have therefore been developed to mitigate the intrinsic limitations of AIMD \cite{behler2007generalized, Behler_2011, Szlachta_2014,  Li_2015, wang2018deepmd, zhang2019active, Jinnouchi_basic}.  
Among them, on-the-fly active learning methods can update the MLFF by training with additional DFT calculations when facing structures of low certainty that have not been sufficiently sampled during the training steps \cite{Li_2015_onthefly, Patra_2017, Jacobsen_2018,  Jinnouchi_basic, Junnouchi_2019_perv, Vandermause2020}. 

The enormous recent progress in the area of MLMD has already had an impact in the SSIC community, e.g., several recent studies employed MLMD to investigate ion-diffusion properties and the role of the host lattice \cite{MIWA2021115567, huang2021deep, staacke2022tackling, gigli2023mechanism,winter2023simulations, Xu2023}. The common target of several studies is to investigate ion-diffusion phenomena on large timescales and to exceed the structural complexity beyond classical bulk systems. Notable examples include MLFFs derived from a self-learning and adaptive database approach, which predicted anisotropic motion of Li ions in Li$_{10}$GeP$_2$S$_{12}$ (LGPS) \cite{MIWA2021115567}. Xu \textit{et al.} employed moment tensor potentials for Li conductors and identified spontaneous polyanion rotation over the timescale of \si{\mu\s}, which were however found to hamper Li-ion migration \cite{Xu2023}. Ref.~\citenum{gigli2023mechanism} showed activation of PS$_4$ flipping events to drive structural transitions to the highly conductive Li$_3$PS$_4$ phase at high temperatures, ruling out paddle-wheel effects in Li-ion conduction via a two-level ML scheme and \si{\ns}-long MLMD simulations. Staacke \textit{et al.} developed a Gaussian approximation potential to tackle ion diffusion within a lithium thiophosphate glass \cite{staacke2022tackling}. These studies exemplify the extent by which structural complexity and/or simulation times of AIMD can be exceeded when using MLMD methods, which is important for obtaining mechanistic insights for SSICs.

Despite these incredible recent efforts, the question of whether and to what extent MLMD simulations can capture the physical mechanism behind ion transport within the strongly anharmonic potential energy surfaces of SSICs remains unaddressed. In particular, when high-throughput screening techniques shall be able to discern materials showing different ion-transport mechanisms, the question of whether MLFFs are accurate across the variety of ion-transport phenomena known to occur in SSICs is very pertinent. Here, an applicability of MLMD can deliver improved microscopic understandings of structure--property relations and the ion-migration mechanism, which will impact the design of novel material candidates. However, such prospects demand an accurate MLMD characterization of both the cation and the host lattice dynamics in SSICs.

In this work, we systematically investigate the dynamical properties of three characteristic SSICs candidates representing different transport mechanisms: AgI, a strongly disordered SSIC showing liquid-like Ag-conduction; Li$_{10}$GeP$_2$S$_{12}$, showing concerted Li$^+$ migration; and Na$_3$SbS$_4$, a Na$^+$ vacancy conductor. These materials represent a variety of challenges for MLFFs including anharmonicities, highly dynamical potential energy surfaces as well as point defects. MLMD simulations were performed and compared with AIMD reference results in terms of the structural and vibrational properties of the SSIC host lattice as well as the diffusion mechanism of the mobile cations. We observe an accurate prediction of the vibrational density of states and diffusion coefficients for all systems. Moreover, even anharmonic distributions in the tilting angles within the host lattice are precisely captured. Most significantly, we demonstrate that MLMD can accurately predict the conduction mechanism of the different systems even in the presence of such extreme lattice anharmonicities. Overall, our findings demonstrate the high suitability of MLMD simulations for broad investigations of SSICs. 

\section{Computational methods}
\label{sec:method}

\textit{Ab initio} molecular dynamics (AIMD) and machine-learning molecular dynamics (MLMD) simulations were carried out with the Vienna Ab initio Simulation Package (VASP) \cite{kresse1996efficient,hafner2008ab}. The projector augmented wave method \cite{blochl1994projector,kresse1999ultrasoft} and the Perdew-Burke-Ernzerhof (PBE) exchange-correlation functional were used for AIMD simulations \cite{perdew1996generalized}. The Brillouin zone was sampled at the $\Gamma$-point only. A canonical ($NVT$) ensemble was applied in each simulation.

\begin{figure}[h!]\centering
    \centering
    \includegraphics[width=.8\textwidth]{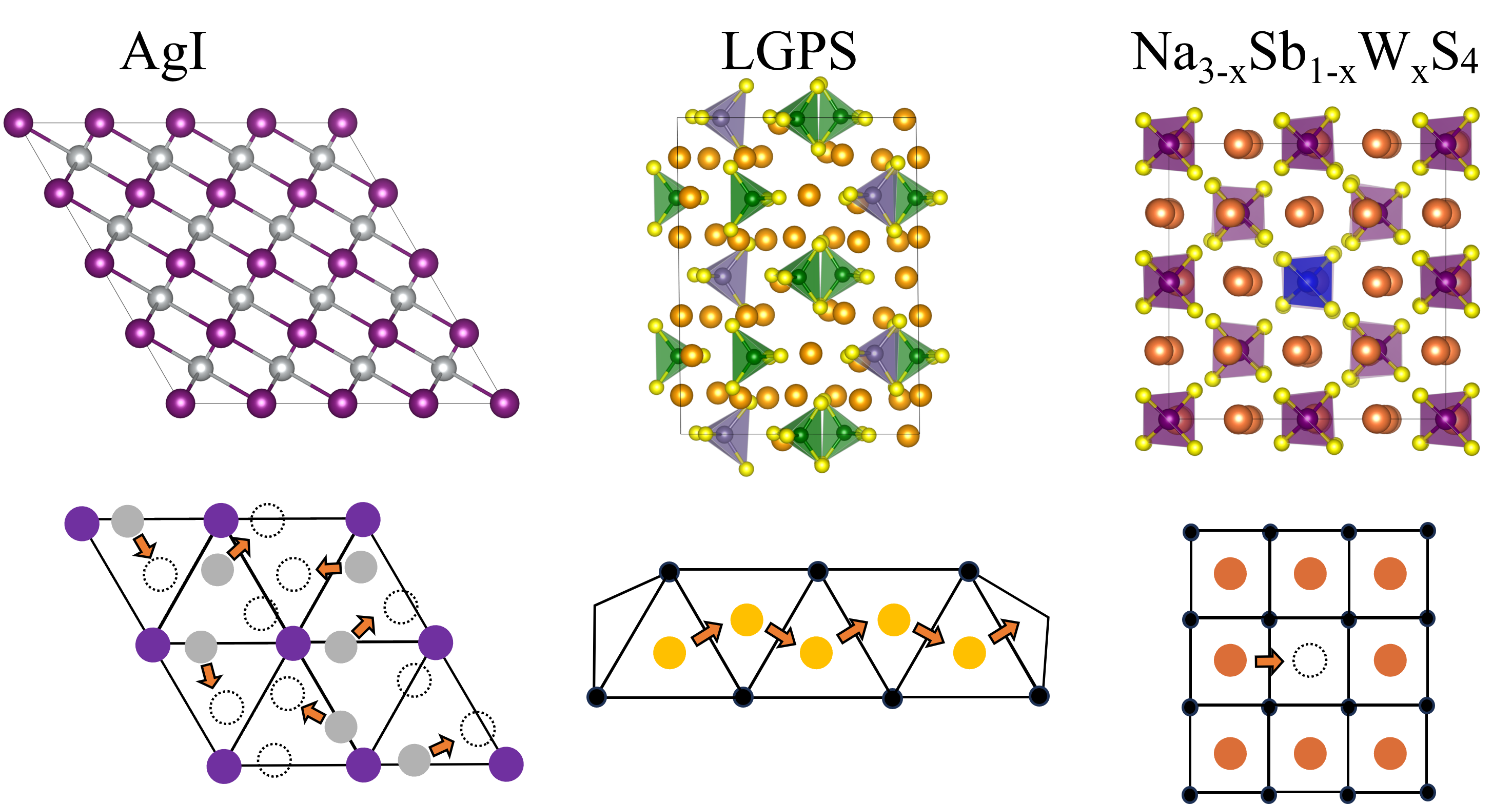}
    \caption{Schematic representation of the solid-state ionic conductors investigated in this work: (Left) AgI, with Ag in silver and I in purple; (mid) Li$_{10}$GeP$_2$S$_{12}$ (LGPS) with Li in bright orange, Ge in grey, P in green and S in yellow; (right) Na$_{3-x}$Sb$_{1-x}$W$_x$S$_4$ with Na in dark orange, W in blue, Sb in purple and S in yellow. The ion-conduction mechanisms of each system are sketched in the lower panels.}
    \label{fig:structures}
\end{figure}

In this work, we train MLFFs 'on-the-fly' using the machine-learning (ML) method implemented in VASP \cite{Junnouchi_2019_perv, Jinnouchi_basic,jinnouchi2020descriptors}.
After convergence of the force field, MLMD simulations are performed using the generated MLFFs to investigate their suitability for the prediction of the atomic dynamics in SSICs. The descriptors of the utilized ML method are based on a non-linear mixing of two- and three-body atomic functions describing the radial and angular distributions, respectively, within the local environment. The implemented method is a kernel-based ML model that uses a Bayesian linear regression to train on the DFT-calculated energy, forces, and stress tensors. The MLMD then makes use of a weighted sum of Kernel similarity measures, with weights obtained from the Bayesian linear regression model, to yield the prediction of forces.
A summary of the ML method is given in the Supporting Information; for further details, we refer the reader to the original literature \cite{Jinnouchi_basic,Junnouchi_2019_perv, jinnouchi2020descriptors}. System-specific parameters for all investigated SSICs (see \cref{fig:structures}) used in the AIMD and MLMD simulations are summarized in the Supporting Information.

We analyze the AIMD and MLFF data in two ways: first, we report the accuracy of the derived MLFFs using the Bayesian estimated error of the force (BEEF), as implemented in VASP \cite{Jinnouchi_basic, Junnouchi_2019_perv, jinnouchi2020descriptors}, which gives us insights into the expected forces and allows to check for convergence of the MLFFs. Additionally, forces from MLFFs are explicitly compared with DFT single-point calculations for randomly selected snapshots along the AIMD trajectories in order to quantify deviations with the root mean square error (RMSE). Hereby, we assess the quality of the MLFFs to describe the forces acting on each element separately.

Second, we calculate a variety of dynamic structural observables that are relevant for SSICs and ion conduction, starting with the atomic pair correlation function, $g(r)$, which quantifies the short- and long-range correlation between two atomic species A and B defined as
\begin{equation}
    g(r) = \frac{1}{N_\textrm{A}N_\mathrm{B}}\sum_{i\in \textrm{A}}\sum_{j\in\textrm{B}}\frac{\delta(|\mathbf{r}_i- \mathbf{r}_{j}| - r)}{4\pi r^2}\,,
\end{equation}
where $N_\textrm{A}$ and $N_\textrm{B}$ are the total number of A and B atoms in the supercell. Given the short-range nature of the descriptors used in the MLFF method that is applied here, an accurate prediction of long-range behavior is an important criterion for their applicability to SSICs and will be assessed via $g(r)$.

The vibrational density of states (VDOS) quantifies the number of modes per unit frequency range and is relevant for ion conduction because, e.g., the zero-frequency component is related to the diffusion constant, $D$, \cite{nitzan2006chemical} and a presence of low-frequency vibrational modes was discussed to imply higher ion conductivity \cite{krauskopf2018comparing, bernges2022considering}.
It is computed as the Fourier transform of the normalized velocity autocorrelation function (VACF):
\begin{equation}
    \textrm{VDOS}(\omega) = \frac{N}{V}\mathcal{F}\left(\frac{\mathrm{VACF}(t)}{\mathrm{VACF}(0)}\right)\,,
    \label{eq_vdos}
\end{equation}
with $N$ and $V$ being the number of contributing atoms and the simulation volume, respectively, and the VACF being defined as
\begin{equation}
    \textrm{VACF}(t) = \langle \mathbf{v}_i(t+\tau)\cdot\mathbf{v}_i(\tau) \rangle_i \,,
\end{equation}
with initial time $\tau$ and $\mathbf{v}_i(t)$ being the velocity of atom $i$ at time $t$. All VDOS spectra are smoothened using a Savitzky–Golay filter with a window of \SI{15}{\per\cm} for improved visibility.

The band center of a given atomic species $i$, $\omega_{\textrm{avg},i}$, is calculated as the average VDOS of species $i$, indicated as a subscript, given as
\begin{equation}
    \omega_{\textrm{avg},i} = \frac{\int \textrm{d}\omega \textrm{VDOS}_i(\omega)\cdot\omega}{\int \textrm{d}\omega \textrm{VDOS}_i(\omega)}\,.
    \label{eq_wavg}
\end{equation}

For AgI, previous work identified that significant angular fluctuations of AgI$_4$ units are involved in anharmonic contributions to the structural dynamics that impact ion conductivity \cite{Brenner_2020}. LGPS and W-doped Na$_3$SbS$_4$ host lattices are also composed of tetrahedral units AS$_4$ (A = Ge, P, Sb, W), whose internal dynamics may impact the motion of mobile cations.
Therefore, we calculate tetrahedral angles in AgI$_4$ (AS$_4$) units representing I--I--I (S--S--S) bond angles, $\theta_i$, via 
\begin{equation}
    \theta_i = \arccos{\left(\frac{\mathbf{d}_{ij}\cdot\mathbf{d}_{ik}}{\abs{\mathbf{d}_{ij}}\abs{\mathbf{d}_{ik}}}\right)}\,,
\end{equation}
where $\mathbf{d}_{ij}$ are atomic distances between iodide (sulfur) ions $i$ and $j$ within a tetrahedron. The angular velocity of tetrahedron $i$ is approximated as 
\begin{equation}
    \omega_{i}(t) = \frac{\theta_i(t+dt)-\theta_i(t-dt)}{2dt}\,,
    \label{eq_anglevel}
\end{equation}
where $dt$ is the MD simulation time step. The time evolutions of $\theta$ and $\omega$ were smoothed using a Fourier filter with a threshold frequency of $\SI{5}{cm^{-1}}$ to retain only the low-frequency dynamics.

Focusing on the mobile ionic species and diffusion properties, we obtain the mean-squared displacement (MSD) according to
\begin{equation}
    \textrm{MSD}(t) = \left\langle \left(\mathbf{r}_i(t)-\mathbf{r}_i(0)\right)^2\right\rangle_i\,,
\end{equation}
where $\langle\cdots\rangle_i$ denotes the average over all cations $i$, and $\mathbf{r}_i(t)$ gives the position of cation $i$ at time $t$. $D$ was then obtained from the MSD at final time $t_\textrm{max}$ as
\begin{equation}
    D = \frac{\textrm{MSD}(t_\textrm{max})}{6 t_\textrm{max}}\,.
    \label{eq_msd}
\end{equation}

Finally, we calculate the van Hove correlation \cite{van1954correlations} and exploit it to quantify temporal correlations between cation dynamics. Specifically, the distinctive part, $G_d$, describes the radial distribution of cations (indexed with $j$) with respect to a reference cation of the same (indexed with $i$) as a function of a time interval $\Delta t$:
\begin{equation}
    G_d(r,\Delta t) = \frac{1}{4\pi r^2N} \left\langle \sum_{i=1}^N\sum_{j=1;j\neq i}^N \delta(r-\abs{\mathbf{r}_i(t+\Delta t)-\mathbf{r}_j(t)}) \right\rangle_t \,,
\end{equation}
with the Dirac delta function $\delta$, the number of cations $N$, and the time average $\langle\cdots\rangle_t$. Conversely, the self part of the van Hove correlation, $G_s$, provides temporal correlations of the radial distribution of cation $i$ with itself as a function of a time interval $\Delta t$:
\begin{equation}
    G_s(r, \Delta t) =\frac{1}{4\pi r^2N} \left\langle \sum_{i=1}^N \delta\left(r-\abs{\mathbf{r}_i(t+\Delta t)-\mathbf{r}_i(t)}\right) \right\rangle_t \,. 
\end{equation}
The relatively low computational costs of MLMD calculations allowed us to conduct five independent runs for each system in order to quantify statistical deviations of selected observables.

\section{Results}
\label{sec:results}

\subsection{Silver iodide (AgI)} 
We start our analysis with silver iodide in its $\alpha$-phase, AgI, which is a prototype solid-state superionic conductor: at temperatures above 420 K, AgI forms a superionic cubic phase with excellent Ag$^+$ conduction. I$^-$ ions form a body-centered cubic (bcc) lattice with a large number of interstices (6 octahedral, 12 tetrahedral, and 24 trigonal), allowing a facile motion of Ag$^+$ ions \cite{Hanson;1975, Stephen;Hull_2004}. The multitude of available Ag$^+$ sites and their strong diffusivity induce a static and dynamic disorder, and the harmonic phonon approximation is breaking down in AgI with important ramification for its ion conductivity \cite{Niedziela2019, Brenner_2020, Ding_2020}. 
All of these render AgI an ideal, challenging system to understand the accuracy and limitations of MLFF to predict the properties and elucidate the conduction mechanisms of SSICs.

\begin{figure}[t]\centering
\includegraphics[width=\textwidth]{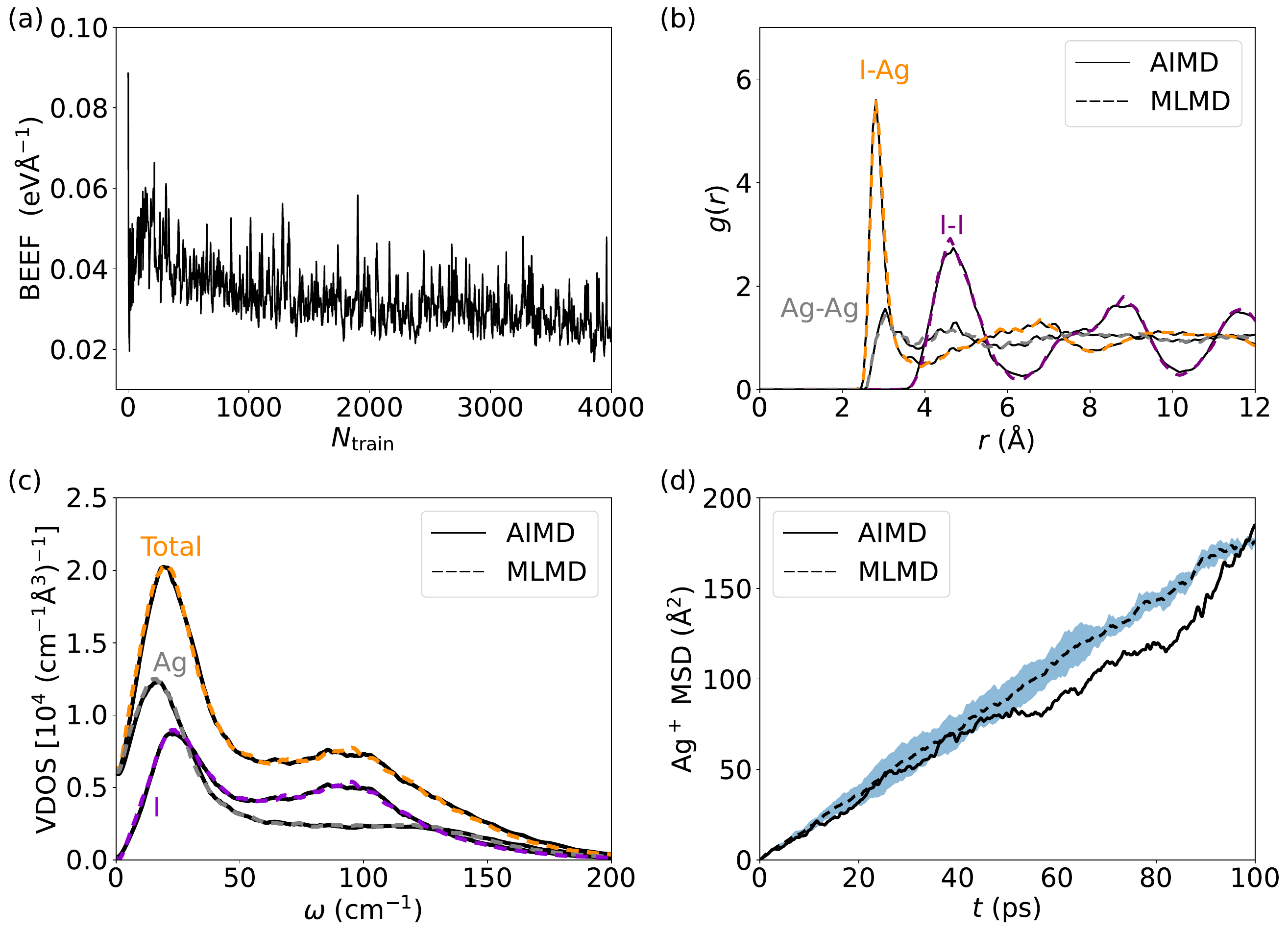} 
\caption{(a) Bayesian error estimate of forces (BEEF) as function of the number of training steps, $N_\textrm{train}$, during training of the MLFF for AgI. Panels (b) and (c) visualize the atomic pair correlation function, $g(r)$, and the vibrational density of states (VDOS), respectively. In panel (d) the mean square displacement (MSD) of Ag$^+$ ions is shown as a function of time, $t$, with the dashed line being the average of 5 MLMD runs and the blue-shaded region the standard deviation.}
\label{fig:AgI_basics}
\end{figure}

We first investigate the MLFF accuracy via the Bayesian error estimate of forces (BEEF) during the training period (\cref{fig:AgI_basics}a), showing well-converged force predictions to $\leq \SI{0.05}{\eV\per\angstrom}$. 
The actual root mean squared errors (RMSE) of the forces in the MLFF, referred to explicit DFT force calculations, are \SI{45.5}{\meV\per\angstrom} and \SI{40.2}{\meV\per\angstrom} for I and Ag atoms, respectively, see Figure~S1, Supporting Information.
Next, we consider the AIMD-computed (see Methods section for computational details) atomic pair correlation function, $g(r)$, which shows a distinct I--Ag peak at \SI{2.8}{\angstrom} and a significant broadening at larger distances, see \cref{fig:AgI_basics}b. I$^-$ ions exhibit long-range order, while the Ag--Ag g(r) shows a liquid-like behavior with a value of $\sim$1 across the whole range. 
Remarkably, MLMD simulations reproduce the AIMD-calculated $g(r)$ of both the iodide host lattice and the disordered Ag$^+$ ions. 
In view of the very different dynamic characteristics of the mobile Ag cation and I host lattice, this finding is important and encouraging.
We further investigate the VDOS (see \cref{fig:AgI_basics}c and Methods section for computational details) and find two broad peaks around 25 and \SI{100}{\per\cm}.
Contributions from Ag$^+$ and I$^-$ ions dominate the former, while the latter is dominated by I$^-$ modes, in good agreement with previous reports \cite{OSullivan_1991, Brenner_2020}. We further obtain non-zero values at \SI{0}{\per\cm} due to slow Ag$^+$ vibrational modes, which is in line with the high ion conductivity of AgI. Our MLMD simulations show an excellent agreement over the whole frequency range from 0 to \SI{200}{\per\cm}, suggesting that MLFFs can accurately capture and reproduce not only structural but also vibrational properties even in the highly complex potential energy surface of AgI. 

We now focus on the Ag$^+$ ion conduction mechanism and investigate the MSD of Ag$^+$ ions, see \cref{fig:AgI_basics}d and Methods section for computational details. We observe a close to linear relationship for MSD$(t)$ in both AIMD and MLMD simulations. Notably, the linear increase of MSD with time is known from Brownian motion,\cite{uhlenbeck1930theory} which together with $g(r)\sim 1$ suggests random dynamics of Ag$^{+}$ ions in a liquid-like environment. 
$D_{\textrm{Ag}^+}$ values extracted from the MSD from AIMD and MLMD simulations are \SI{2.61e-5}{\cm\squared\per\s} and \SI{2.99e-5}{\cm\squared\per\s}, respectively, which are in reasonable agreement and in line with previous studies \cite{Kvist_1970, Shommers_1977}. 

\begin{figure}[t]\centering
\includegraphics[width=\textwidth]{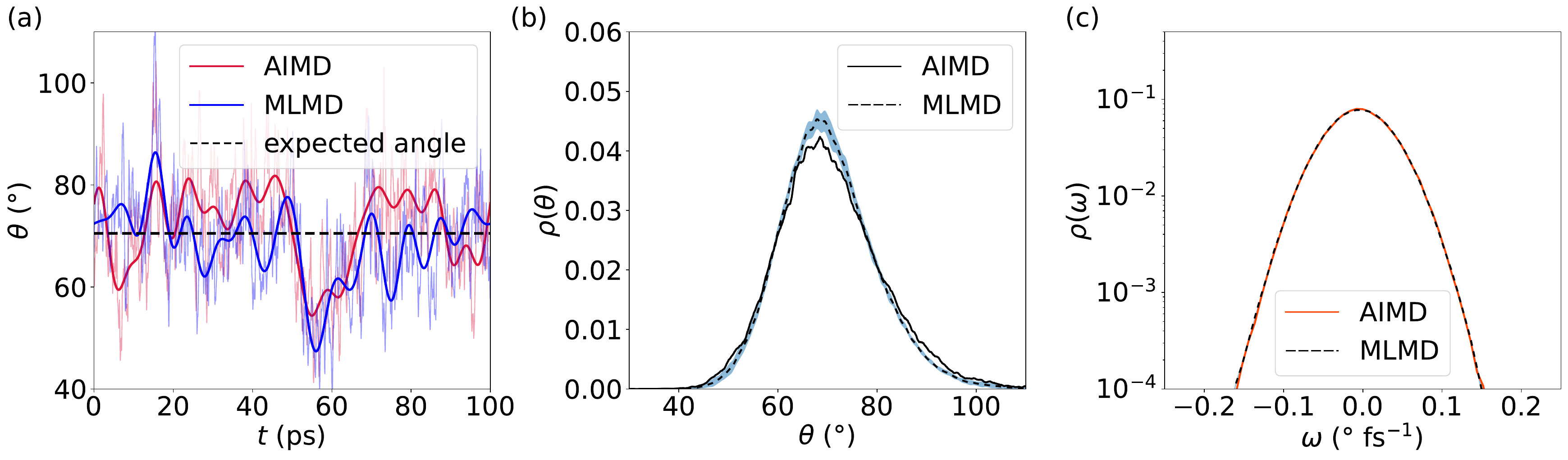} 
\caption{(a) Representative time evolution of I--I--I bond angle, $\theta$, for selected AgI$_4$ tetrahedra (see Figure~S2, Supporting Information, for schematic representation). Thick (thin) lines in panel (a) represent the Fourier filtered (unfiltered) values of $\theta$ using a cutoff frequency of \SI{5}{\per\cm}. (b) Probability density, $\rho(\theta)$, as obtained from 48 tetrahedra within the AgI supercell.
(c) Probability densities of the angular velocity, $\rho(\omega)$, calculated from the time evolution of $\theta$. For $\rho(\theta)$ and $\rho(\omega)$, data averaged from 5 independent MLMD runs are shown and the light-blue area indicates the standard deviation.}
\label{fig:AgI_angle}
\end{figure}

Because anharmonic relaxational dynamics of the iodide host-lattice were discussed as a driving force for Ag$^+$ conduction \cite{Brenner_2020}, it is important to examine whether MLFFs can capture such dynamic effects. First, in both AIMD and MLMD we observe large deviations of I--I--I bond angles, $\theta$, from the expected value of \SI{70.5}{\degree} to below \SI{45}{\degree} and above \SI{100}{\degree}, see \cref{fig:AgI_angle}a. As shown by Brenner et al. \cite{Brenner_2020}, these distortions remain present for long timescales, as also seen in the slow relaxational components of the $\theta$ angles we compute (see \cref{fig:AgI_angle}a). Interestingly, AIMD and MLMD simulations show a comparable extent of deviations from the expected value. A statistical analysis of the angle probability density, $\rho(\theta)$, taken from 48 different tetrahedral angles in the supercell along the trajectory, shows that values between \SI{30}{\degree} and \SI{110}{\degree} occur in the simulations (see \cref{fig:AgI_angle}b). $\rho(\theta)$ as computed in AIMD shows a more skewed form, while the counterpart from MLMD is slightly more pronounced at \SI{70.5}{\degree}. The tail at small values of $\theta$ is captured well by MLMD but a more pronounced disagreement is observed for larger values of $\theta$. The standard deviation among several MLMD runs is found to be small, which indicates a high degree of reproducibility of these simulations for capturing such distinct local distortions. 

\begin{figure}[t]\centering
\includegraphics[width=\textwidth]{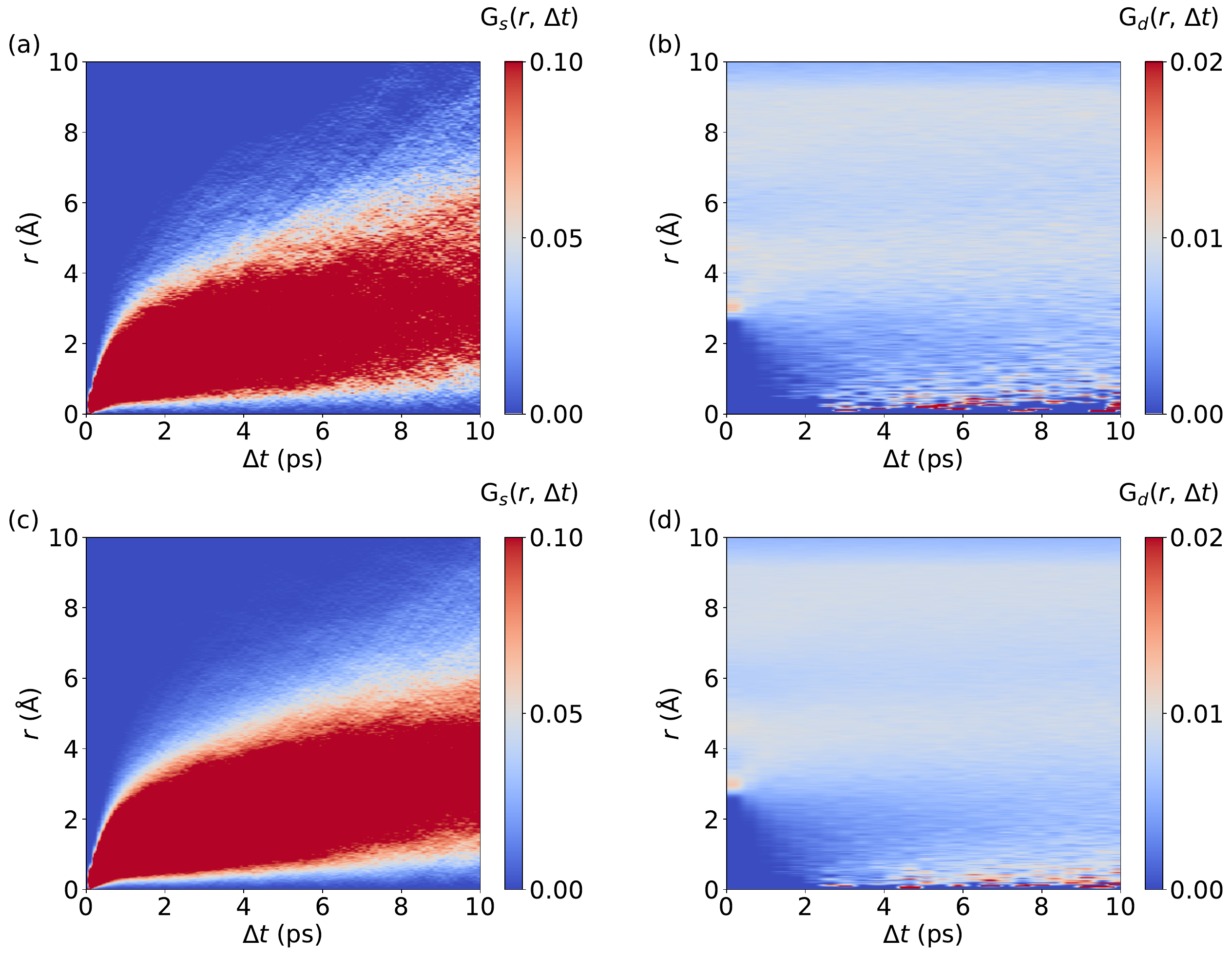} 
\caption{Spatiotemporal correlations of Ag$^+$ ion migration in AgI: (a) and (c) visualize the self-part, $G_s(r, \Delta t)$, of the van Hove correlation extracted from AIMD and MLMD simulations, respectively. Panels (b) and (d) visualize the distinctive part, $G_d(r, \Delta t)$, from AIMD and MLMD simulations, respectively. MLMD data were averaged across 5 independent trajectories.}
\label{fig:AgI_vanHove}
\end{figure}

To further examine the $\theta$ distributions, we compute the angular velocity, $\omega$, see \cref{eq_anglevel}. Because of the highly dynamic nature of AgI, the restoring force acting on iodide ions are anticipated to be rather small, \textit{i.e.} small $\omega$ values may be predicted. This is in line with the results shown in \cref{fig:AgI_angle}a, where deviations in $\theta$ from the expected value occur for long timescales of $\sim\SI{20}{ps}$. Indeed, we find small values of $\omega$ between \SI{-0.2}{\degree\per\fs} and \SI{0.2}{\degree\per\fs} (see \cref{fig:AgI_angle}c), confirming the weak restoring forces. MLMD simulations accurately capture the AIMD probability density of angular velocities, $\rho(\omega)$, over the entire range suggesting an accurate prediction of the dynamical changes in the iodide host lattice (see \cref{fig:AgI_angle}c). 

A potential reason for the deviations in the $\theta$ distribution of MLMD compared to AIMD may be rooted in the limited phase space that is being sampled during MLFF training. To investigate this hypothesis, we retrained the MLFF at an increased temperature of \SI{800}{\kelvin} and performed another MLMD simulation at \SI{500}{\kelvin} using the alternative MLFF. In this way, the system explores a larger part of phase space during MLFF training, expanding into strongly anharmonic regions due to the increased thermal fluctuations. However, the deviation in $\rho(\theta)$ is found to be similar even when training is conducted at higher temperatures (Figure~S3, Supporting Information). Consequently, we may assign the remaining deviations to the choice of two- and three-body descriptors in the MLFFs, for which more complex ones may be required to capture all aspects of $\rho(\theta)$ with a high accuracy.

Finally, we investigate the van Hove correlation function of Ag$^+$ ions, considering the self part, $G_s$ and the distinctive part, $G_d$, see \cref{fig:AgI_vanHove} and Methods section for the computational details. In line with the monotonous increase of the MSD (cf. \cref{fig:AgI_basics}d), an increasing width in the $G_s$ values with time is observed. Interestingly, $G_s$ values at $r=\SI{0}{\angstrom}$ quickly drop to zero as $\Delta t$ is increased. This implies that Ag$^+$ ions barely return to their previous location, which is a futher aspect confirming the randomness in the Ag$^+$ ion dynamics.
Furthermore, after fast broadening of the strongly correlated $G_s$ domain within $\Delta t\leq \SI{0.5}{\pico\second}$, its center increases continuously with time. The diffusive character of Ag$^+$ dynamics is visible as a rapid and continuous increase of $G_s$ values that reflect traveled distances of Ag$^+$ ions, contributing to the remarkable ion conductivity of AgI. At the same time, no significant $G_d$ peaks are observed. Rather, the $G_d$ data confirm the fully homogeneous distribution of cations over space and time without significant temporal correlations between different Ag$^+$ ions. Importantly, MLMD simulations accurately capture the shape and intensities of $G_s$ and $G_d$ as obtained in AIMD.
Overall, our results suggest that MLMD is able to capture the structural, vibrational, diffusive, and angular properties of AgI even with the intricate potential energy surface this material experiences at higher temperatures.

\subsection{Li$_{10}$GeP$_2$S$_{12}$ (LGPS)}

Lithium-based SSICs emerged as a promising alternative to replace their liquid counterparts without changing the full battery stack because of their excellent ion conductivities \cite{kim2021solid}. Among them, Li$_{10}$GeP$_2$S$_{12}$ (LGPS) was discovered as one of the first Li-based SSICs with an ion conductivity comparable to those of liquid electrolytes \cite{HONG1978117, Kamaya2011, Adams2012, Kuhn2013, Kuhn2013_2, Kwon2015, Kato2016, SUN2016798, He2017,KANNO200097,Kyuki2020,Ye_2020, MIWA2021115567}. Many studies have investigated the underlying mechanism that drives Li$^+$ migration through the solid host lattice \cite{Kanno_2001, Kato2016, NACHIMUTHU2022101223, Wang2022}, and AIMD studies suggested a concerted migration of Li$^+$ ions to be at play \cite{Xu2012, He2017}. Specifically, the coupled motion of multiple Li$^+$ ions can lower migration energy barriers significantly compared to a scenario where ions migrate individually. This mechanism of coupled Li$^+$ motion is similar to the situation in liquid electrolytes \cite{Donati1998, Keys2011} and likely is accompanied by anharmonic dynamical changes in the local potential energy surface of the material. It is critically important to assess whether MLFFs can capture such concerted dynamic phenomena, despite being based on local two- and three-body descriptors and a limited size of the AIMD snapshots available for MLFF training. Consequently, we investigate the accuracy of MLMD simulations to capture the concerted motion of Li$^+$ ions as well as the interactions with the host lattice in the prototypical LGPS material.

\begin{figure}[t]\centering
\includegraphics[width=\textwidth]{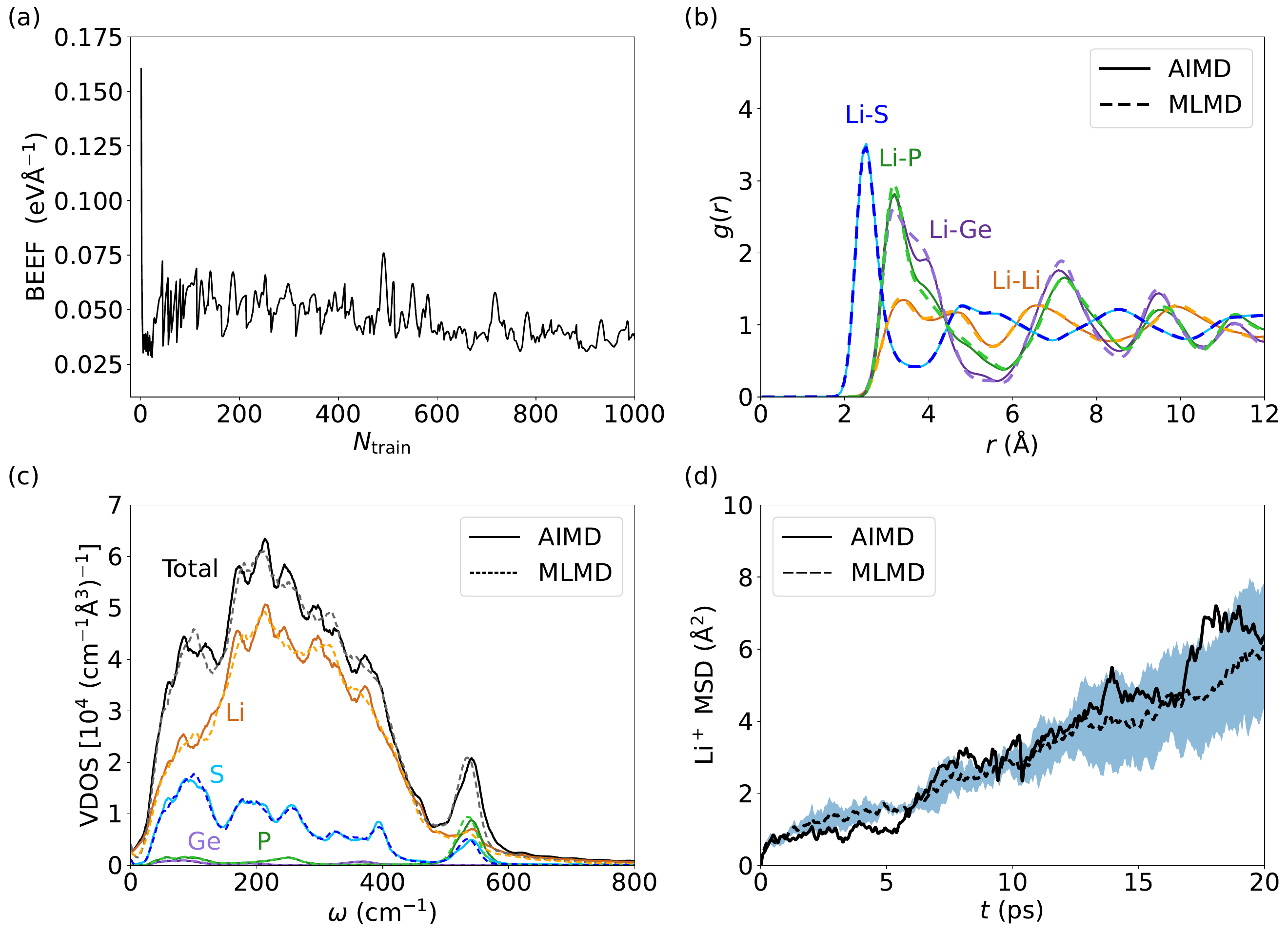}
\caption{(a) BEEF as function of $N_\textrm{train}$ during training of the MLFF for LGPS. Panels (b) and (c) visualize $g(r)$ and the VDOS, respectively. In panel (d) the MSD of Li$^+$ ions is shown as a function of $t$, with the dashed line being the average of 5 MLMD runs and the blue-shaded region the standard deviation.}
\label{fig:LGPS_PCF}
\end{figure}

As above, we analyze the accuracy of the trained MLFF first via the force errors during the training period (\cref{fig:AgI_basics}a) and observe fast convergence of the BEEF to values below \SI{0.05}{\eV\per\angstrom} (see \cref{fig:LGPS_PCF}a). For the individual chemical species, we observe RMSE values of MLFF with respect to DFT force values of \SI{38.7}{\meV\per\angstrom} for Li, \SI{87.2}{\meV\per\angstrom} for Ge, \SI{98.9}{\meV\per\angstrom} for P, and \SI{56.1}{\meV\per\angstrom} for S (see Figure~S4, Supporting Information).
The calculated $g(r)$ show a first-order peak around \SI{2.5}{\angstrom} for Li--S and broadened higher-order peaks (see \cref{fig:LGPS_PCF}b). The $g(r)$ data for Li--P and Li--Ge both show a peak at \SI{3.5}{\angstrom} and exhibit an overall structure that is consistent with the expected long-range order of the host lattice. By contrast, a suppressed long-range order is found for Li--Li with $g(r)\approx1$ across the whole range. MLMD simulations accurately capture the features of $g(r)$ for Li with most elements, in particular for Li--Li and Li--S. However, we do observe differences in the Li--P and Li--Ge $g(r)$ peak intensities, whereas their peak positions are accurately captured. Specifically, their first and second $g(r)$ peaks are not precisely reproduced by MLMD. 
These deviations are in line with the enhanced RMSE values for Ge and P atoms, which is likely caused by the small amount of Ge and P atoms in the supercell, \textit{i.e.} 8 and 16 atoms, respectively, and consequently the low number of local reference configurations for each type, as mentioned in the computational details.
It is noted that MLMD further captures the $g(r)$ within the polyanions of the host lattice (see Figure~S5, Supporting Information). 

We now investigate the VDOS of LGPS (see \cref{fig:LGPS_PCF}c) and obtain good agreement between AIMD and MLMD simulations for the entire frequency range. All element-wise contributions, particularly the ones of Li$^+$ ions, are closely mirrored in MLMD. Li vibrations appear throughout the whole frequency range up to \SI{650}{\per\cm}, dominating in particular the region between \SIrange{200}{400}{\per\cm}, whereas contributions by Ge and P are found mostly for higher-frequency modes at \SI{550}{\per\cm}. There are some deviations between AIMD and MLMD visible in the high-frequency range of the polyanion data, which are likely not very relevant because ion migration is driven by lower-frequency modes. Furthermore, we observe a finite VDOS value at zero frequency, pointing to the Li$^+$ migration. 
MLMD simulations, averaged across 5 runs, show a continuous increase in the MSD (\cref{fig:LGPS_PCF}d). In contrast to our findings for AgI, we notice that the variation in the Li$^+$ MSD is substantial, which suggests an increased stochastic nature of the ion-migration mechanism. The MSD from the single AIMD run follows the MLMD one to a good extent. $D_{\textrm{Li}^+}$ values of \SI{5.33e-5}{\cm\squared\per\s} and \SI{4.27e-5}{\cm\squared\per\s} from AIMD and MLMD simulations, respectively, were extracted, which are in line with previous computational studies \cite{Mo_2011}. We assign the observed deviations to the increased stochastic character of Li migration in LGPS. Thus, large simulation times -- exceeding feasible AIMD timescales -- are required to predict reliable MSD values. 

\begin{figure}[t]\centering
\includegraphics[width=\textwidth]{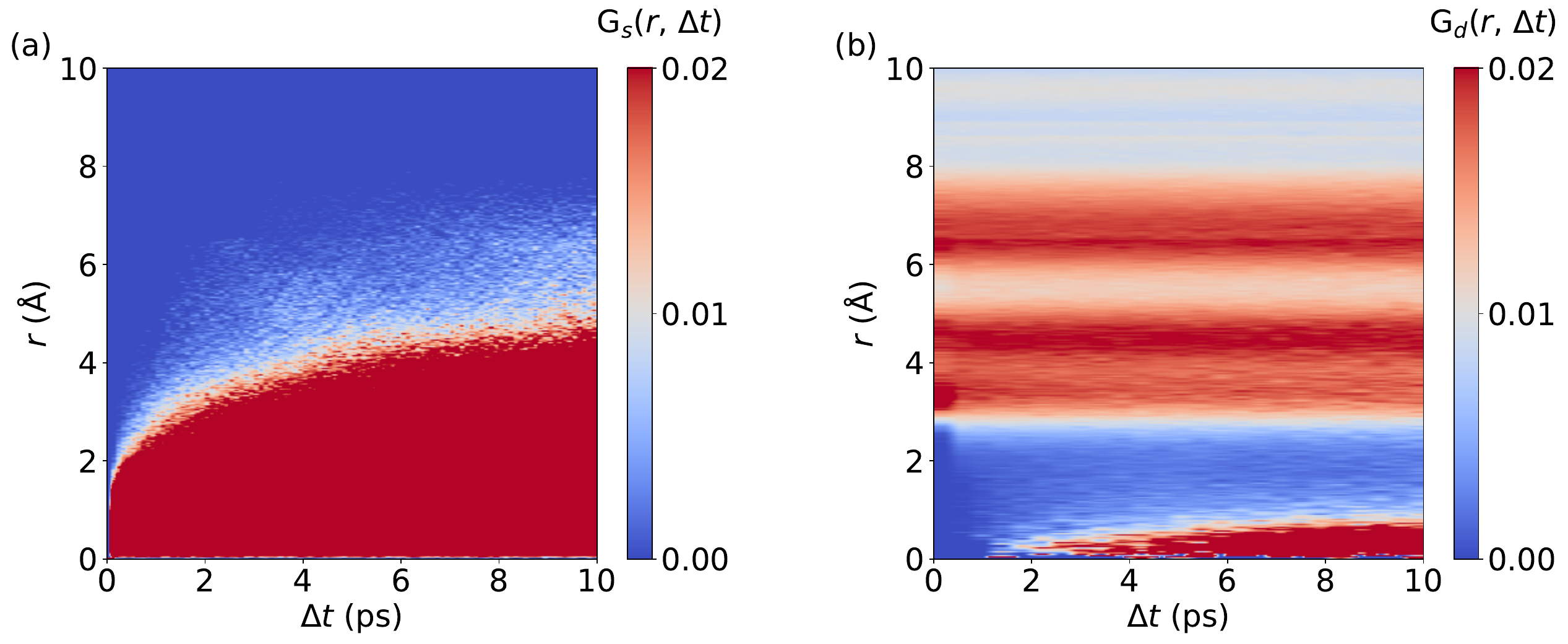}
\caption{Spatiotemporal correlations of Li$^+$ ion migration in LGPS: Panels (a) and (b) visualize $G_s(r, \Delta t)$ and $G_d(r, \Delta t)$, respectively. MLMD data were averaged across 5 independent trajectories.
}
\label{fig:LGPS_VHC}
\end{figure}

We now analyze the migration mechanism of Li$^+$ ions. Here, we calculate the self and distinctive part of the van Hove correlation function for Li$^+$ ions in LGPS from AIMD and MLMD trajectories. MLMD results of $G_d(r, \Delta t)$ show a substantial correlation at \SI{0}{\angstrom} starting at $\sim\SI{1}{\ps}$. This means that a Li$^+$ ion follows a migrating one within timescales of around \SI{1}{\ps}, supporting the concerted motion of Li$^+$ ions (see \cref{fig:LGPS_VHC} and Figure~S6, Supporting Information). Strong correlations exist around \SI{4}{\angstrom} and \SI{6.5}{\angstrom}, confirming that the long-range order of Li$^+$ ions remains unaffected even though Li$^+$ ions migrate, in line with previous results \cite{He2017}. $G_s(r, \Delta t)$ shows a continuous increase in correlation during the simulations, showing that the Li$^+$ ions are migrating, which is consistent with the MSD (see \cref{fig:LGPS_PCF}d). In contrast to AgI, we observe a substantial correlation at $r=\SI{0}{\angstrom}$ for all $\Delta t$ values. This underpins that a substantial amount of Li$^+$ ions remain located near their equilibrium positions, while all Ag$^+$ ions contribute to the overall conduction in AgI.
Notably, the MLFF shows a fair agreement for $G_s(r, \Delta t)$ and $G_d(r, \Delta t)$ with the AIMD reference (see Figure~S6, Supporting Information). Deviations in the overall values and intensities are caused by the limited available timescales from AIMD simulations. Importantly, we can still confirm that MLMD captures the concerted migration mechanism in LGPS.

\begin{figure}[t]\centering
\includegraphics[width=\textwidth]{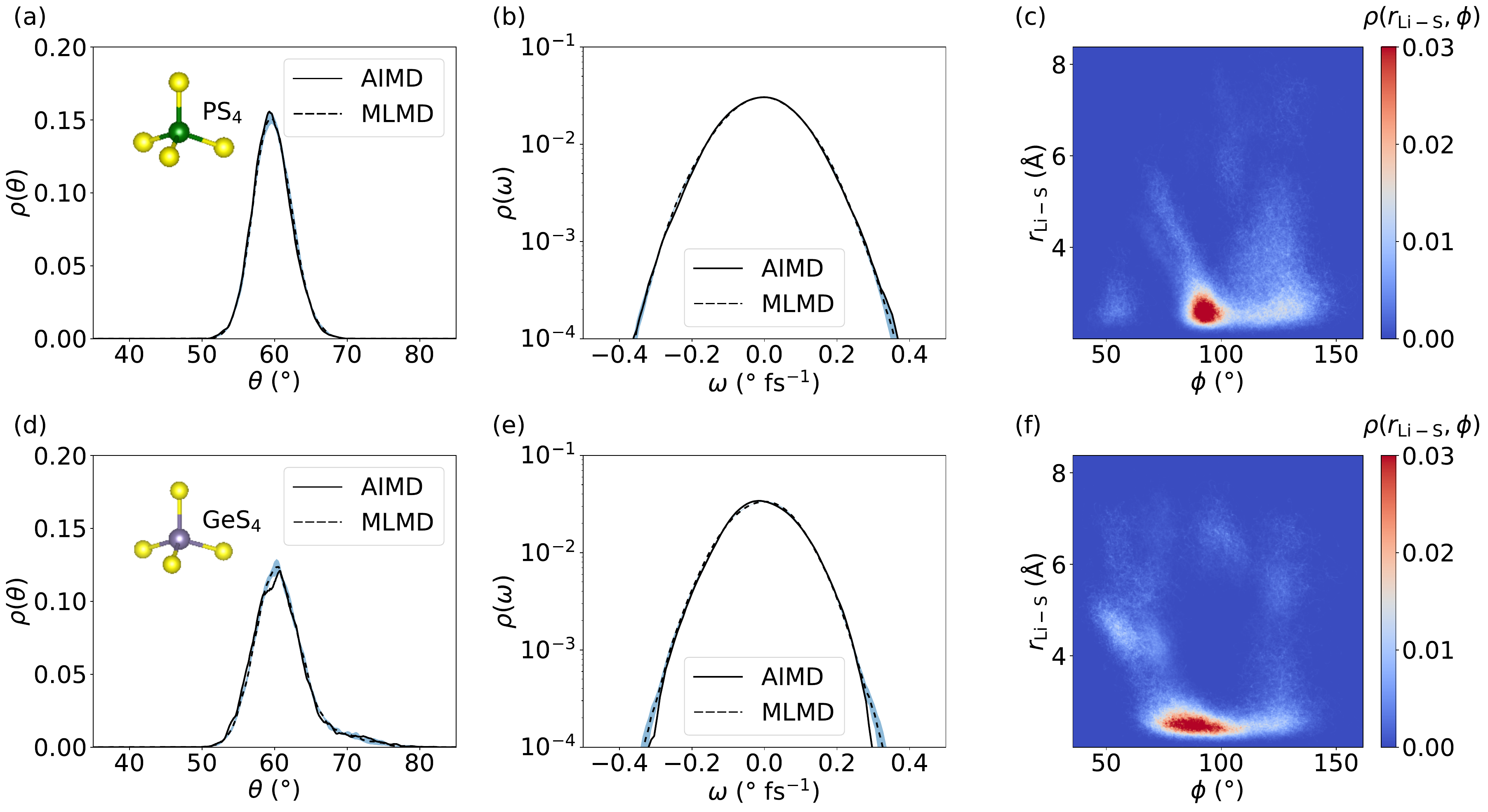}
\caption{
Probability density, $\rho(\theta)$, of $\theta$ angles in (a) PS$_4$ and (d) GeS$_4$ tetrahedra in LGPS. Panels (b) and (e) show $\rho(\omega)$ for PS$_4$ and GeS$_4$, respectively. Panels (c) and (f) depict the 2D probability density, $\rho(r_\textrm{Li--S}, \phi)$, quantifying correlations between the PS$_4$ and GeS$_4$ units, respectively, and Li--S bond lengths $r_\textrm{Li--S}$. All MLMD data were averaged across 5 independent runs and the light-blue area indicates the standard deviation.
}
\label{fig:LGPS_angle}
\end{figure}

We now focus on the dynamics of the PS$_4$ and GeS$_4$ tetrahedra in the LGPS host lattice because distortions in the polyanion backbone are likely to play a crucial role in facilitating cation diffusion \cite{Jun2022}. Moreover, it is crucial to investigate the aforementioned consequences of the increased RMSE of the MLFF for Ge and P, as well as the deviations seen in $g(r)$, for the dynamical properties of the host lattice. 
We compute the distribution of $\theta$ angles within the PS$_4$ and GeS$_4$ tetrahedra and compare the AIMD and MLMD probability densities, $\rho(\theta)$, see \cref{fig:LGPS_angle}. The agreement between AIMD and MLMD is overall good. We generally find small variations in the $\theta$ angles suggesting relatively rigid tetrahedra. PS$_4$ tetrahedra show fluctuations that are largely harmonic in the range between \SI{50}{\degree} and \SI{70}{\degree}. Interestingly, the angles in GeS$_4$ tetrahedra exhibit a distribution that is significantly more asymmetric with an extended tail toward large values. This suggests the presence of substantial anharmonicities in the rotational dynamics of GeS$_4$ tetrahedra. The angular velocity distribution, $\rho(\omega)$, further underlines the accuracy of the MLMD simulations in capturing not only the angular distribution but also their dynamical changes (\cref{fig:LGPS_angle}). The higher RMSE values of the forces from MLFFs for P and Ge atoms (see Figure~S4, Supporting Information) may explain the remaining deviations present in the angular distributions. In contrast, the RMSE of Li forces is considerably lower, in line with our finding that Li$^+$ migration is captured well.

Having assessed the dynamical properties of mobile cations and host lattice separately, we now investigate the interaction between PS$_4$ and GeS$_4$ polyanions and adjacent Li$^+$ ions. An accurate description from MLMD is relevant for capturing pertinent dynamic coupling phenomena, such as paddle-wheel effects, which potentially drive Li$^+$ ion migration \cite{Jansen_1991, zhang2022exploiting, Xu2023}. We compute the 2D probability distribution, $\rho(r_\textrm{Li--S}, \phi)$, which measures occurrences of Li--S distances, $r_\textrm{Li--S}$, as a function of P--S--Li bond angles, $\phi$, see also Figure~S7, Supporting Information, for a graphical representation. 
Highest correlation between PS$_4$ tetrahedra and Li ions is found in a small range of $\SI{80}{\degree}\leq\phi\leq\SI{95}{\degree}$ at small Li--S distances of $\leq\SI{3}{\angstrom}$ (see \cref{fig:LGPS_angle} and Figure~S8, Supporting Information). Two less intense branches are found leaving this highly correlated region. The first one reaches higher $\phi$ angles of up to \SI{130}{\degree} with Li--S distances spread over the whole considered range. The second, less-intense branch simultaneously goes to smaller $\phi$ angles and larger Li--S distances. The correlation of Li--S distances with GeS$_4$ tetrahedra shows a squeezed form with small changes in the distances, showing $\phi$ angles of \SIrange{75}{110}{\degree}. This suggests a larger rotational degree of freedom of Li ions around GeS$_4$ than around PS$_4$ tetrahedra, which we may assign to the larger mass and less negative charge of the [GeS$_4$]$^{2-}$ polyanions. Interestingly, Li--S distances increase only when $\phi\leq\SI{70}{\degree}$ or $\phi\geq\SI{120}{\degree}$, representing a potential indicator for the design of host lattice structures to enable Li ion migration. Direct comparison with AIMD results (see Figure~S8, Supporting Information) shows an excellent match in the probability distributions with MLMD data, supporting the accurate prediction of interactions between mobile cations and the host lattice. These results underline that MLMD captures and reproduces the structural, vibrational, and concerted nature of Li ion migration in LGPS. 

\subsection{Sodium Thioantimonate (Na$_3$SbS$_4$)}

Na-based SSICs gained significant attention as a potential replacement for their Li-based counterparts. Sodium thiophosphates, Na$_3$PS$_4$ \cite{hayashi2012superionic,de2016diffusion}, and sodium thioantimoniates, Na$_3$SbS$_4$ \cite{wang2016air, zhang2016vacancy, hayashi2019sodium}, have been the focus of many studies in this context because of their integrability into all-solid-state sodium batteries and due to their high ion conductivities \cite{banerjee2016na3sbs4, zhang2018synthesis}. Moreover, a variety of interesting findings have been reported on the atomic dynamics in these materials. This includes vibrational anharmonicity \cite{gupta2021fast, brenner2022anharmonic}, a rich polymorphism \cite{famprikis2019new, famprikis2021insights}, and coupled dynamics of anions and cations that have been discussed in the context of paddle-wheel effects \cite{famprikis2019new, zhang2019coupled}. A peculiarity of current Na-based SSICs is their tunable ion conductivity by introducing Na vacancies via aliovalent doping. Several studies suggested that Na$^+$ diffusion in pristine materials remains absent even at temperatures as high as \SI{900}{\kelvin} and only becomes viable upon introduction of defects \cite{bo2016computational,moon2018vacancy}. Such can be achieved through a variety of different dopants, including halides \cite{chu2016room, feng2019studies,jalem2022theoretical} and tungsten \cite{fuchs2019defect, hayashi2019sodium, feng2021heavily}, which commonly substitute a pnictogen or chalcogen, respectively, and leave Na$^+$ vacancies behind to counterbalance the change in oxidation state. Na$^+$ ions may then diffuse via vacancy migration with DFT-computed migration barriers as low as \SI{0.1}{eV}, resulting in ion conductivities in the \SI{}{\milli\siemens\per\cm} range \cite{rush2017unraveling}.

AIMD has been instrumental in demonstrating the migration of Na$^+$ vacancies within the network of [PnCh$_4$]$^{3-}$ (Pn=\{P, Sb\}; Ch=\{S, Se\}) polyanions \cite{moon2018vacancy, famprikis2019new, gupta2021fast, jalem2022theoretical}. Specifically, the anharmonic character of the host-lattice vibrations and local symmetry breaking, which cannot be captured by harmonic phonon calculations alone, have motivated use of finite-temperature MD simulations \cite{krauskopf2018local, krauskopf2018comparing, famprikis2021insights, brenner2022anharmonic}. However, computational costs associated with doped scenarios of, e.g., Na$_3$PnCh$_4$ SSICs, are very high. First, large supercells are needed to mimic small doping concentrations at the Pn site. Second, Na$^+$ migration occurs by hopping across vacancies, which are rare events despite modest ion conductivities and low migration barriers found in these materials. Consequently, long simulation times are needed to sample sufficiently many hopping events and describe the coupling between host lattice and migrating Na$^+$ ions. 

MLMD trained via AIMD data may well constitute a promising way to overcome these computational challenges, potentially allowing for the study of extended length and time scales, which we investigate here for the scenario of aliovalent doping. Specifically, we focus on tungsten-doped Na$_{3-x}$Sb$_{1-x}$W$_x$S$_4$, which was shown to attain excellent ion conductivities \cite{fuchs2019defect, feng2021heavily}. The mechanism involves W substituting a Sb site, forming [WS$_4$]$^{2-}$ polyanions that exhibit different charges than the original [SbS$_4$]$^{3-}$ ones, and a variation in the chemical environment of S atoms. As a consequence, Na vacancies, V$_\textrm{Na}^-$, are introduced which may diffuse within the crystal. The local variation in chemical bonds and charge as well as changes in the potential energy surface due to V$_\textrm{Na}^-$ migration, can be expected to result in challenges for the training procedure of MLFFs. 

We start by considering the accuracy of MLMD simulations for pristine Na$_3$SbS$_4$ at $T=\SI{300}{\kelvin}$. The MLFF quickly converges within 1000 steps, showing maximum force errors of below \SI{0.02}{\eV\per\angstrom} after 200 steps already (see Figure~S9, Supporting Information). This results in precise predictions of $g(r)$ among all elements in the anion host lattice and the Na$^+$ ions (see Figure~S10, Supporting Information). The VDOS shows the expected polyanion vibrations at larger frequencies of \SIrange{350}{400}{\per\cm}, and broad contributions in the range of $\leq\SI{200}{\per\cm}$ caused in particular by Na and S vibration (see Figure~S11, Supporting Information). The latter suggests the presence of anion--cation coupling, in line with previous studies \cite{famprikis2021insights, gupta2021fast, brenner2022anharmonic}. Notably, the MLMD data capture mode frequencies across the entire frequency range of vibrations, but moderate differences in their intensities remain present. We further observe that Na$^+$ ions merely oscillate around their lowest-energy positions, while migration throughout the pristine lattice remains absent in our AIMD and MLMD simulations (see Figures~S12--S14, Supporting Information), as previously reported \cite{bo2016computational}.

\begin{figure}[t]\centering
\includegraphics[width=\textwidth]{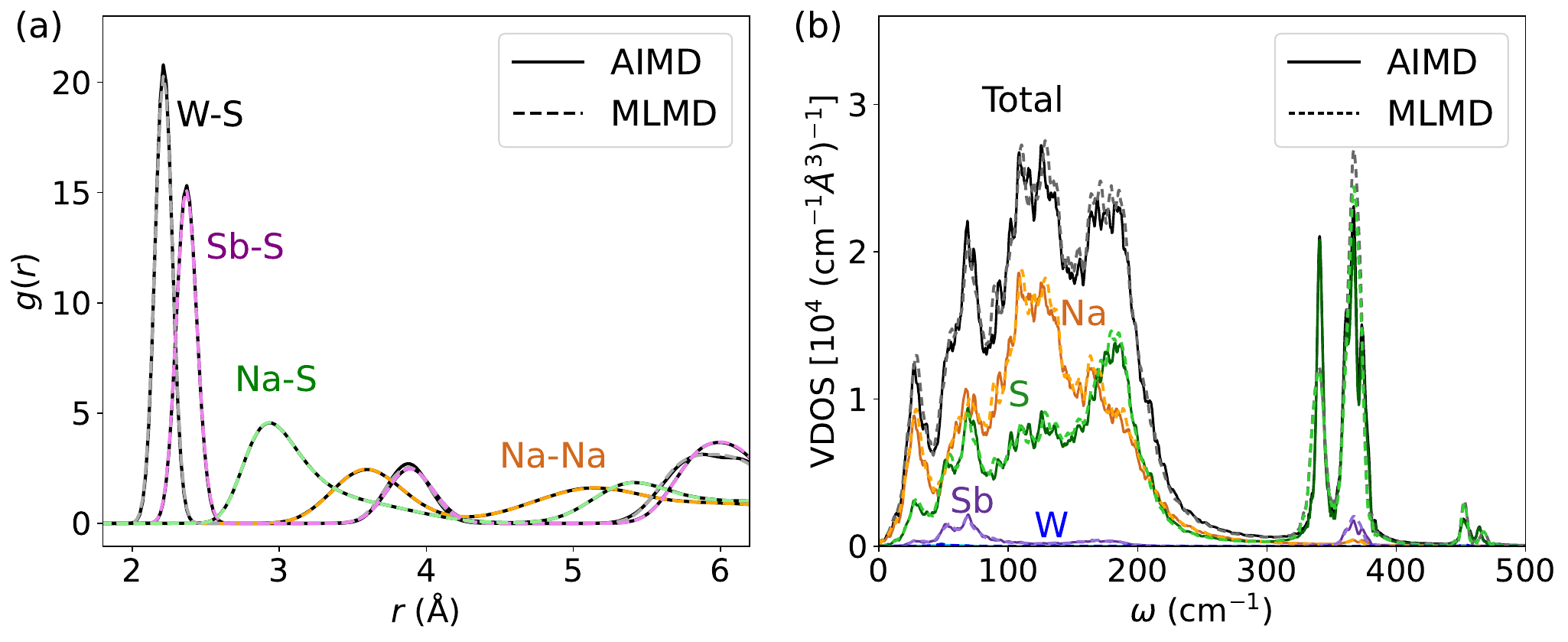}
\caption{(a) $g(r)$ and (b) VDOS for W-doped Na$_{2.94}$Sb$_{0.94}$W$_{0.06}$S$_4$ as computed from AIMD and MLMD. 
}
\label{fig:NSS_PCF}
\end{figure}

Next, we investigate the accuracy of MLFFs for the tungsten-doped Na$_{2.94}$Sb$_{0.94}$W$_{0.06}$S$_4$. A new set of force fields is trained starting from the MLFF of the pristine Na$_3$SbS$_4$. The BEEF values during the training procedure largely remain below \SI{0.02}{\eV\per\angstrom} after around 1000 steps (see Figure~S15, Supporting Information). As expected, the error of the MLFF is higher than the one for the pristine system due to increased variations in the local coordination of S atoms in the W-substituted polyanions. Explicit calculation of errors from MLFF to DFT forces were \SI{14.0}{\meV\per\angstrom}, \SI{41.4}{\meV\per\angstrom}, \SI{24.3}{\meV\per\angstrom} and \SI{52.3}{\meV\per\angstrom} for Na, Sb, S, and W, respectively (see FIgure~S16, Supporting Information). Comparing these results to the above findings, these forces are in the range of what we found for AgI and below the ones reported for Ge and P in LGPS. Consequently, the MLMD simulations accurately describe $g(r)$ within the [SbS$_4$]$^{3-}$ polyanions as well as between Na$^+$ and S$^{2-}$ ions when compared to AIMD data (see \cref{fig:NSS_PCF}a). We observe only minor deviations in the W--S $g(r)$, where MLMD results show slightly lower peak intensities than the AIMD ones (see also Figure~S17, Supporting Information). The W--S $g(r)$ shows a pronounced peak at \SI{2.21}{\angstrom}, confirming that bond lengths are shortened compared to Sb--S of \SI{2.36}{\angstrom} in the pristine case because of the more positive W$^{6+}$ ions. $g(r)$ for Na--Na remains largely unaffected by doping and still shows a first-order peak at \SI{3.61}{\angstrom} with broadened long-range distributions. 

An important question is whether MLMD simulations accurately describe the vibrational properties of W-doped Na$_{2.94}$Sb$_{0.94}$W$_{0.06}$S$_4$. Indeed, we observe good agreement between MLMD and AIMD data for the overall shape of the VDOS and individual element-wise contributions in the region $\leq\SI{300}{\per\cm}$ (see \cref{fig:NSS_PCF}b) despite substantial changes in the local chemical environments. However, the intensity of higher-energy sulfide modes shows larger deviations despite their frequencies being captured well, as explicitly shown in $\Delta$VDOS plots (Figure~S18, Supporting Information). While an accurate representation of low-frequency modes is more important for describing SSICs, reducing such deviations for the high-frequency vibrational regime could be achieved with improved MLFFs. We also investigate the Na band center, given by the average vibrational frequencies of the projected Na VDOS (see \cref{eq_wavg}), because it was recently identified as a good indicator for changes in the overall lattice stiffness \cite{krauskopf2018comparing}. Here, AIMD and MLMD simulations give Na band center values of $\omega_\textrm{avg,Na}=\SI{127.5}{\per\cm}$ and \SI{126.5}{\per\cm}, respectively, showing excellent agreement. By contrast, the pristine Na$_3$SbS$_4$ exhibits a larger value of \SI{129.7}{\per\cm} (\SI{129.1}{\per\cm}) from MLMD (AIMD) simulations, which shows that a small softening of the lattice occurs upon introducing W dopants and that this correlates with the higher Na$^+$ conductivity.

\begin{figure}[t]\centering
\includegraphics[width=\textwidth]{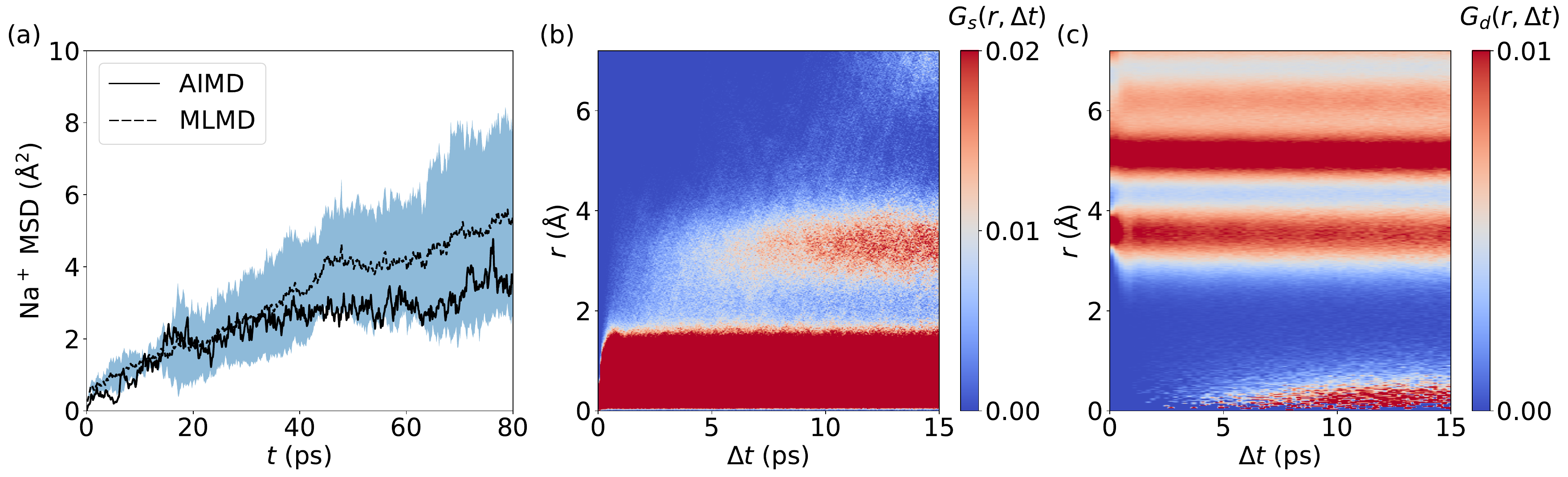}
\caption{(a) MSD of Na$^+$ ions in Na$_{2.94}$Sb$_{0.94}$W$_{0.06}$S$_4$ as a function of $t$, with the dashed line being the average of 5 MLMD runs and the blue-shaded region the standard deviation.
Spatiotemporal correlations of Na$^+$ ion migration: Panels (b) and (c) visualize $G_s(r, \Delta t)$ and $G_d(r, \Delta t)$, respectively. MLMD data were averaged across 5 independent trajectories.}
\label{fig:NSS_MSD_vanhove}
\end{figure}

The prediction of Na$^+$ diffusion coefficients from AIMD simulations is greatly complicated by random Na$^+$ hopping processes. Consequently, long simulations and/or large cell sizes are required to obtain reliable MSD values, which again underlines the need for reliable MLMD. Our analysis of the Na$^+$ MSD from MLMD simulations highlights these complications by showing a wide spread of MSD values over time (\cref{fig:NSS_MSD_vanhove}a), from which we obtain $D_\textrm{Na}^+=(1.13\pm0.69)\times\SI{e-6}{\cm\squared\per\s}$. The MSD computed from the single AIMD trajectory lies within the standard deviation of MLMD values. Van Hove correlations for Na$_{2.94}$Sb$_{0.94}$W$_{0.06}$S$_4$ exhibit a behavior that significantly deviates from AgI and LGPS: $G_s$ develops an additional region of strong correlations after \SI{7}{\ps} at approximately \SI{3.5}{\angstrom}, corresponding to the distance between neighboring Na$^+$ sites and confirming the migration of Na$^+$ ions (see \cref{fig:NSS_MSD_vanhove}b and Figure~S19, Supporting Information). Notably, the region between the two strongly correlated $G_s$ features is only weakly populated, which points to the migration of Na$^+$ ions via thermally activated hopping. Concerted migration can be ruled out from $G_d$ (see \cref{fig:NSS_MSD_vanhove}c), showing only a slowly increasing component after \SI{6}{\ps}.  

\begin{figure}[t]\centering
\includegraphics[width=\textwidth]{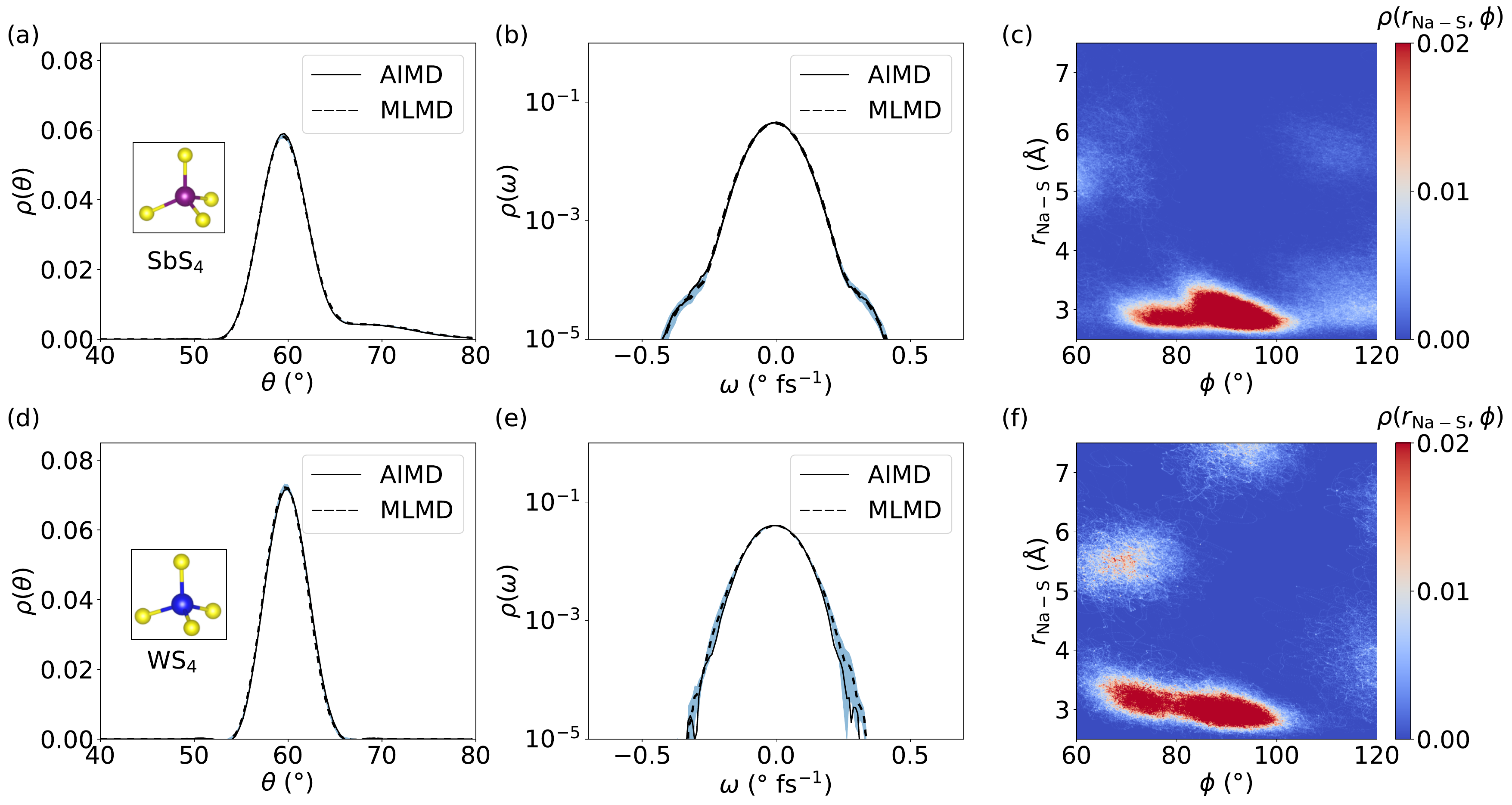}
\caption{Probability density, $\rho(\theta)$, of $\theta$ angles in (a) SbS$_4$ and (d) WS$_4$ tetrahedra in Na$_{3-x}$Sb$_{1-x}$W$_{x}$S$_4$. Panels (b) and (e) show $\rho(\omega)$ for SbS$_4$ and WS$_4$, respectively. Panels (c) and (f) depict the 2D probability density, $\rho(r_\textrm{Na--S}, \phi)$, quantifying correlations between the Sb--S--Na and W--S--Na angle $\phi$, respectively, and  Na--S bond lengths $r_\textrm{Na--S}$. All MLMD data were averaged across 5 independent runs and the light-blue area indicates the standard deviation.}
\label{fig:NSS_angle}
\end{figure}

Because several studies suggested that coupling between cations and host lattice is a key factor in facilitating cation diffusion \cite{famprikis2019new, zhang2019coupled}, we investigate the dynamics of SbS$_4$ and WS$_4$ tetrahedra in the host lattice of Na$_{2.94}$Sb$_{0.94}$W$_{0.06}$S$_4$.
Distributions of $\theta$ angles within SbS$_4$ and WS$_4$ tetrahedra are obtained in order to compare the AIMD and MLMD probability densities, $\rho(\theta)$, see \cref{fig:NSS_angle}a and d, respectively. The overall agreement between AIMD and MLMD is again satisfactory. The overall small dynamic changes of $\theta$  indicate that the tetrahedra form relatively rigid structures. Notably, $\theta$ of SbS$_4$ tetrahedra exhibit an asymmetric distribution with an extended tail toward larger angle values whereas $\theta$ of WS$_4$ tetrahedra exhibit an essentially symmetric characteristic.
This indicates the existence of anharmonicity in the rotational motions of SbS$_4$ tetrahedra. This is also visible in the angular distribution, $\rho(\omega)$, of SbS$_4$ tetrahedra, which shows small deviations around \SI{-0.5}{} and \SI{0.5}{\degree\per\fs} from harmonic behavior. Notably, $\rho(\omega)$ of WS$_4$ tetrahedra shows a more harmonic distribution. The overall small deviations between AIMD and MLMD are found to be more pronounced for WS$_4$, which likely is a result of the small sampling size of WS$_4$. 

We now focus on correlations between SbS$_4$ and WS$_4$ tetrahedra and neighboring Na$^+$ ions. We obtain the 2D probability density $\rho(r_{\mathrm{Na-S}}, \phi)$, by defining the distance, $r_{\mathrm{Na-S}}$, between neighboring Na$^+$ ions and S atoms, and the dihedral angle $\phi$, see \cref{fig:NSS_angle} for MLMD and Figure~S20, Supporting Information, for AIMD results. The largest correlation between SbS$_4$ tetrahedra and Na ions exists in a small range of $\SI{80}{\degree}\leq\phi\leq\SI{100}{\degree}$ with $r_{\textrm{Na-S}}$ remaining below $\leq\SI{4}{\angstrom}$. Minor intensities are found in two regions: one in a range of $\SI{100}{\degree}\leq\phi\leq\SI{120}{\degree}$ and $r_{\mathrm{Na-S}}$ of $\leq\SI{4}{\angstrom}$, and the other in $\SI{60}{\degree}\leq\phi\leq\SI{80}{\degree}$ and $\SI{5}{\angstrom}\leq r_{\mathrm{Na-S}}\leq\SI{6}{\angstrom}$. 
The noticeable intensities at lower $r_{\mathrm{Na-S}}$ values showing a broadened $\phi$ distribution between \SI{80}{\degree} and \SI{120}{\degree}indicate that Na$^+$ ions experience a significant degree of rotational freedom around SbS$_4$ tetrahedra.
In regard to WS$_4$ tetrahedra, the intensity in the range $\SI{60}{\degree}\leq\phi\leq\SI{80}{\degree}$ and $\SI{5}{\angstrom}\leq r_{\mathrm{Na-S}}\leq\SI{6}{\angstrom}$ is found to be more pronounced than in case of SbS$_4$ ones. This observation suggests a facilitated ion migration when Na$^+$ cations are passing by WS$_4$ tetrahedra, which can be explained by W--S bond lengths being shorter than the Sb--S ones and the fact that WS$_4$ tetrahedra occupy less space. Consequently, Na$^+$ ions are given more space to migrate in the proximity of the W-dopant. The absence of correlated branches as we found in case of LGPS (see \cref{fig:LGPS_angle}) can be rationalized by the nature of Na$^+$ migration discussed above when investigating $G_s$ for Na$_{2.94}$Sb$_{0.94}$W$_{0.06}$S$_4$ (see \cref{fig:NSS_MSD_vanhove}): Na$^+$ cations spend comparably little time in-between equilibrium sites, which is characteristic of a thermally-activated cation hopping, whereas Li$^+$ ions in LGPS spend more time in-between their equilibrium sites.
Altogether, these findings show that the host lattice dynamics of Na$_{2.94}$Sb$_{0.94}$W$_{0.06}$S$_4$ are correlated with cation diffusion and suggest that ion conduction can be tuned through adjusting the dynamics of the host lattice.

\section{Discussion}
\label{sec:discussion}

Our systematic analysis found that MLMD simulations are accurate for the vibrational properties of mobile cations and the host lattice. Anharmonic properties are predicted well, and the interactions between mobile cations and the polyanionic backbone are precisely captured for all investigated systems. These results are especially intriguing as various ion-conduction mechanisms were investigated: liquid-like behavior of Ag$^+$ ions in AgI, concerted migration of Li$^+$ in LGPS, and thermally activated V$_\textrm{Na}^-$ vacancy hopping in Na$_3$SbS$_4$. To the best of our knowledge, a systematic investigation of MLFF accuracy covering such diverse ion migration phenomena in SSICs has not been reported yet. The majority of existing studies rely on the accuracy of MLFFs in comparison to DFT reference calculations. However, recent reports highlight that benchmarking the accuracy of forces alone does not necessarily align with an accurate prediction of relevant material properties \cite{fu2023forces}. Here, both structural and vibrational properties of mobile cations and host lattice are found to be captured well, providing confidence that the utilized MLFFs are highly suitable for a broad range of SSICs.

We emphasize that such accurate predictions from the exploited MLMD model were not necessarily expected. The locality of descriptors, which only include atoms that are located within a certain cutoff radius, is often discussed as one downside for larger systems next to advantages such as  simplicity and transferability \cite{pronobis2018many, unke2021machine}. Such concerns have led to the development of novel MLFF methods that correct for nonlocal effects including long-range electrostatic interactions \cite{grisafi2019incorporating, ko2021fourth, unke2021spookynet, kabylda2023efficient}. In the case of SSICs, the short-range nature of interactions and the large dynamical distortions exhibited by the ions likely aids the kernel-based ML procedure. A dominance of short-range interactions can lead to long-range phenomena being less relevant effectively, which can rationalize our accurate MLMD simulation results. Correlated ion phenomena at short ranges still remain well represented as we have demonstrated in the case study of concerted migration of Li ions in LGPS. 

Finally, we would like to provide a compact perspective on the relevance of our findings for future development of materials and the role MLMD simulations may play in this regard.
Fast and reliable MLMD simulations with \textit{ab initio} accuracy can open novel possibilities when guiding experimental materials design and characterization. Raman spectroscopy, for example, represents one of the most widely used, non-destructive techniques for providing vibrational fingerprints of a solid-state material. Even though the first approaches towards high-throughput Raman calculations and characterization have been made recently \cite{ cui2019decoding,liang2019high, sheremetyeva2020machine, bagheri2023high, qi2023recent}, first-principles calculations of Raman spectra remain essential for the interpretation of experimental measurements and prediction of vibrational properties in novel material systems. Commonly used Raman calculations based on the harmonic approximation are, however, not sufficient to capture the entire vibrational dynamics of materials that are as anharmonic as the investigated SSICs here. These inherent limitations can be overcome through molecular dynamics simulations that account for the whole anharmonicity of such material \cite{putrino2002anharmonic, thomas2013computing,  ditler2022vibrational}, which in the past found only limited attention due to the tremendous computational costs associated with AIMD. The excellent accuracy of the presented MLMD results for the variety of vibrational properties in the investigated SSICs may render calculations of finite-temperature Raman spectra viable and hereby inspire novel data-driven approaches, combining ML-models for Raman activities \cite{raimbault_etal_2019, sommers_etal_2020, shang_wang_2021, han_etal_2022,  berger2023polarizability, grumet2023delta, lewis_etal_2023} with MLMD to launch a flourishing synergy between experiments and computation. 

Moreover, we envision that computationally efficient and precise MLMD simulations may enrich the physical information that is contained in datasets for a tailored design of novel materials \cite{nolan2018computation, ohno2020materials, merchant2023scaling}. Here, the majority of existing approaches for data-driven screening of suitable material candidates mainly relies on properties that are derived in the ground state \cite{fujimura2013accelerated, wang2015design, sendek2018machine, muy2019high, kahle2020high}. Incorporating vibrational properties as well as information about the ion-conduction mechanism derived from MLMD simulations may improve the selectivity of existing screening approaches and potentially deepen our understanding of the structure--property relations of SSICs.

\section{Conclusion}
\label{sec:conclusion}

In summary, we have provided a systematic investigation of the accuracy of MLMD simulations to predict the dynamical properties associated with ion conduction across various mechanisms in a broad range of SSICs. Results from MLMD were compared to AIMD reference data using several measures that include the actual force errors, structural and dynamical properties of mobile cations as well as of the host lattice, and correlations among the motion of cations and rotational degrees of freedom in the polyanion backbone. Our results demonstrate that precise predictions of dynamical properties of SSICs exhibiting different transport mechanisms that involve liquid-like cation motion in AgI, concerted cation migration in LGPS, and thermally-activated vacancy hopping in W-doped Na$_3$SbS$_4$ are feasible. Through our MLMD simulations, we could further provide a detailed analysis of the anharmonic character of host lattice vibrations, which take a significant role in enabling a facile migration of mobile cations. Our findings are very encouraging and may further enhance trust in MLMD simulation results and their suitability for a broad range of SSICs with different physical mechanisms governing ion migration and structural variability. We believe that the increasing availability of MLMD methods can enhance synergies among experiments, computations, and material design efforts. An exciting avenue along this quest revolves around the development of novel data-driven screening techniques for SSICs, which will be aided by finite-temperature materials predictions through MLMD simulations at relatively low computational cost.

\section*{Conflict of interest}
The authors declare no conflict of interest.

\section*{Supporting Information}
Additional details of the MLFF theory; detailed numerical parameters and settings for AIMD and MLMD simulations; additional figures on the analysis of force error, as well as of structural and dynamical SSIC properties.

\begin{acknowledgements}
Funding provided by the Alexander von Humboldt--Foundation in the framework of the Sofja Kovalevskaja Award, endowed by the German Federal Ministry of Education and Research, by the Deutsche Forschungsgemeinschaft {via} Germany's Excellence Strategy - EXC 2089/1-390776260, and by the TUM-Oerlikon Advanced Manufacturing Institute, are gratefully acknowledged. The authors further acknowledge the Gauss Centre for Supercomputing e.V. for funding this project by providing computing time through the John von Neumann Institute for Computing on the GCS Supercomputer JUWELS at Jülich Supercomputing Centre.
\end{acknowledgements}

\bibliography{references}

\clearpage

\end{document}